# Structural evolution of Ti/Cu multilayers as a function of period thickness

S.S. Sakhonenkov,[a] A.U. Gaisin,[a] A.O. Petrova,[a] V.A. Matveev[b] and E.O. Filatova[a]

[a] *Institute of Physics, Saint Petersburg State University, Ulyanovskaya Str. 1, Peterhof, St. Petersburg 198504, Russia*

[b] *Petersburg Nuclear Physics Institute named after B.P. Konstantinov of National Research Centre «Kurchatov Institute», mkr. Orlova roshcha 1, Gatchina, Leningradskaya Oblast, 188300, Russia*

**Abstract**

Ti/Cu multilayers with periods ranging from 4 to 52.5 nm were synthesized by magnetron sputtering to examine how the period thickness affects morphology, crystallization, texture development, and preservation of periodicity. The structural evolution was analyzed using complementary transmission electron microscopy techniques with X-ray diffraction and reflectometry. The results show that the period thickness governs the balance between interfacial transition-region formation, crystallization, and growth-induced morphological instabilities. At the smallest period of 4 nm, the structure has strongly reduced periodic contrast, which may be attributed to extended interfacial regions, partial Cu–Ti intermixing, and suppression of well-defined layer formation. At 10 nm, the multilayer structure is preserved but remains affected by accumulated roughness and layer waviness; local modification of the Cu lattice is suggested to be significant relative to the bilayer period, likely due to mixed Cu–Ti interfacial regions and/or coherent strain. With increasing period, the multilayers become more regular, and the layers show progressive crystallization and texture development during growth. Cu crystallizes predominantly in the fcc structure with a strong Cu(111) contribution, whereas Ti exhibits a more complex structural evolution, from an amorphous or weakly crystallized state near the substrate toward a more ordered and textured state closer to the surface.

## Introduction

Ti–Cu-based systems, including bulk alloys, thin films, and multilayer coatings, are of interest for a wide range of applications. The combination of titanium and copper enables materials with an attractive set of properties, including good mechanical performance, biocompatibility, corrosion resistance, and relatively low formation temperatures for a number of phases. At the same time, the properties of Ti–Cu systems are strongly dependent on their microstructural state, including phase composition, crystallite size, degree of texture, layer morphology, and the state of interphase boundaries.

The structural state of titanium in thin-film form at the nanoscale is of particular importance. In the bulk state, titanium has a stable α phase with a hexagonal close-packed lattice (hcp) at room temperature and transforms into a β phase with a body-centered cubic lattice (bcc) at elevated temperatures. However, in thin films and nanostructures, titanium can exhibit more complex phase formation behavior. In addition to the α-Ti phase, which is stable at room temperature, the formation of a metastable fcc-like state in thin films has been observed or discussed in several studies [1–3]. The formation of this phase has also been considered for multilayer systems, where the phase stability of Ti may be affected by interfacial energy and coherent stresses [4]. Overall, these results indicate that the microstructure of Ti films with thicknesses ranging from a few nanometers to several hundred nanometers is sensitive to the deposition method and processing parameters, including particle energy, substrate temperature, deposition rate, working-atmosphere composition, and the state of the growth surface. These factors govern the competition between amorphization, crystallite nucleation, texture development, and the accumulation of growth-induced roughness. In multilayer systems, this competition is further complicated by the presence of periodically repeated interphase boundaries, where interdiffusion, the formation of transitional Cu–Ti regions, local stresses, and changes in the crystal structure of individual layers may occur.

Ti/Cu multilayer coatings are of particular interest for neutron optics [5,6]. Periodic or aperiodic sequences of layers with different neutron-optical contrast can be used as supermirrors, extending the angular range of efficient reflection beyond that provided by ordinary total external reflection. In working supermirrors, the period, defined as the total thickness of a pair of layers, usually varies through the coating thickness, allowing the Bragg condition to be satisfied for a broad range of de Broglie wavelengths at a fixed reflection geometry. For such systems, preservation of the designed periodicity, sharp interphase boundaries, and roughness control are critically important, since these parameters determine the reflectivity and stability of the neutron-optical contrast.

Ti/Cu supermirrors are considered a nondepolarizing alternative to conventional NiMo/Ti coatings for polarized-neutron experiments that require both high magnetic fields and precise preservation of neutron-beam polarization during transport through a neutron guide [7]. In NiMo-containing coatings, 5–20% Mo is used to suppress the ferromagnetic properties of nickel [8]; however, residual magnetization still remains and may lead to spin-dependent scattering and beam depolarization, especially in strong magnetic fields [7,9]. Replacing the ferromagnetic component with diamagnetic copper substantially reduces the magnetic contribution and thereby improves polarization stability [10]. It has previously been shown that Ti/Cu mirrors can provide low depolarization at the level required for precision experiments (~$10^{-4}$), while modern m = 2 Ti/Cu supermirrors demonstrate high reflectivity and thermal stability [6,11].

At the same time, the transition to supermirrors with higher m values is inevitably associated with a reduction in the thickness of individual layers in the lower part of the aperiodic structure. In this regime, the contribution of transition regions becomes comparable to the thickness of the main layers; therefore, even moderate interdiffusion, chemical smoothing of the profile, or growth-induced roughness can substantially reduce the neutron-optical contrast. Additional complexity arises from the crystallization of Cu and Ti. On the one hand, crystallite formation and texture development may stabilize layer growth; on the other hand, grain structure and columnar growth can enhance interlayer roughness and distort the coating periodicity.

Various interface-engineering approaches are used to control the structure of interphase boundaries in Ti/Cu systems. In particular, reactive deposition in nitrogen or synthetic-air atmospheres has been used to suppress the growth of large crystallites, reduce roughness, and decrease the spin-dependent contribution to reflection [5,6]. In addition, data on oxidized Cu–Ti multilayer films indicate that the presence of oxygen may limit Cu–Ti interdiffusion [12]. However, the influence of crystallization of the main layers on the evolution of transition regions in Ti/Cu multilayer structures remains insufficiently studied, especially in the range of periods where the thickness of individual layers becomes comparable to the characteristic extent of interphase regions.

The present work aims to establish the relationship between the period of a Ti/Cu multilayer structure, layer morphology, crystallization development, and preservation of periodicity. Periodic Ti/Cu multilayer structures with different nominal periods from 4 to 52.5 nm are studied as model objects, making it possible to separate the basic mechanisms of structural evolution from the additional complexity characteristic of aperiodic working supermirrors. This approach allows us to trace how, with increasing individual layer thickness, the balance changes between transition-region formation, crystallization of Cu and Ti, texture development, and the accumulation of growth-induced morphological inhomogeneities. The obtained results are important for the further optimization of Ti/Cu coatings intended for neutron optics and other applications where controlled interphase structure and stable multilayer periodicity are required.

## Experimental section

All the multilayer systems and reference films were synthesized by magnetron sputtering using the PAPK facility at the NRC Kurchatov Institute – PNPI (Gatchina). To model different local regions of a depth-graded supermirror, a series of multilayer mirrors was fabricated with nominal layer thicknesses varying between 1.00 nm and 26.25 nm. For each multilayer stack, the thicknesses of the Cu and Ti layers within each bilayer were set equal, ensuring a symmetric period. The period is defined as the combined thickness of a pair of alternating layers, $d = d_{Cu} + d_{Ti}$. All samples were fabricated on highly oriented Si(111) substrates. The number of bilayer repetitions (N) was adjusted for each structure to maintain an approximately constant total multilayer stack thickness of 420 nm. This approach minimizes the relative contribution of the substrate to the reflected signal and ensures comparability of the diffraction and reflectivity data across samples. The resulting multilayer systems are denoted as $[Ti/Cu]_N$, where $N$ is the number of periods. In this notation, the layer adjacent to the substrate is written first (left), while the topmost layer is given last (right). Table 1 summarizes the multilayer structures and reference films investigated in this work along with their nominal thicknesses, estimated based on the sputtering rate.

**Table 1.** Nominal parameters of synthesized samples: $N$ – the number of periods, $l$ – thickness of the individual layers, $d$ – period thickness.

| Structure | $N$ | $l$, nm | $d$, nm |
|---|---|---|---|
| Cu | - | 100 | - |
| Ti | - | 100 | - |
| [Ti/Cu] | 105 | 2/2 | 4 |
| [Ti/Cu] | 42 | 5/5 | 10 |
| [Ti/Cu] | 28 | 7.5/7.5 | 15 |
| [Ti/Cu] | 21 | 10/10 | 20 |
| [Ti/Cu] | 14 | 15/15 | 30 |
| [Ti/Cu] | 8 | 26.25/26.25 | 52.5 |

The sputtering system is equipped with planar magnetrons with constant current sources. All multilayer systems and reference films were deposited in an argon atmosphere. To establish the required working atmosphere in the sputtering chamber, an argon gas supply system is employed. Argon is delivered from a high-pressure cylinder (150 atm) through a pressure regulator that reduces the pressure to 1 atm. The process chamber of the deposition system houses three magnetrons. A dual-circuit cooling system is implemented for thermal management of the magnetrons. In the primary circuit, distilled water circulates and is cooled via a heat exchanger by tap water from the secondary circuit. During operation, the temperature at the cathode–target interface of the magnetrons does not exceed 100 °C. The active area of each magnetron (target area) measures 100 mm × 500 mm. The substrate-to-target distance is fixed at 75 mm.

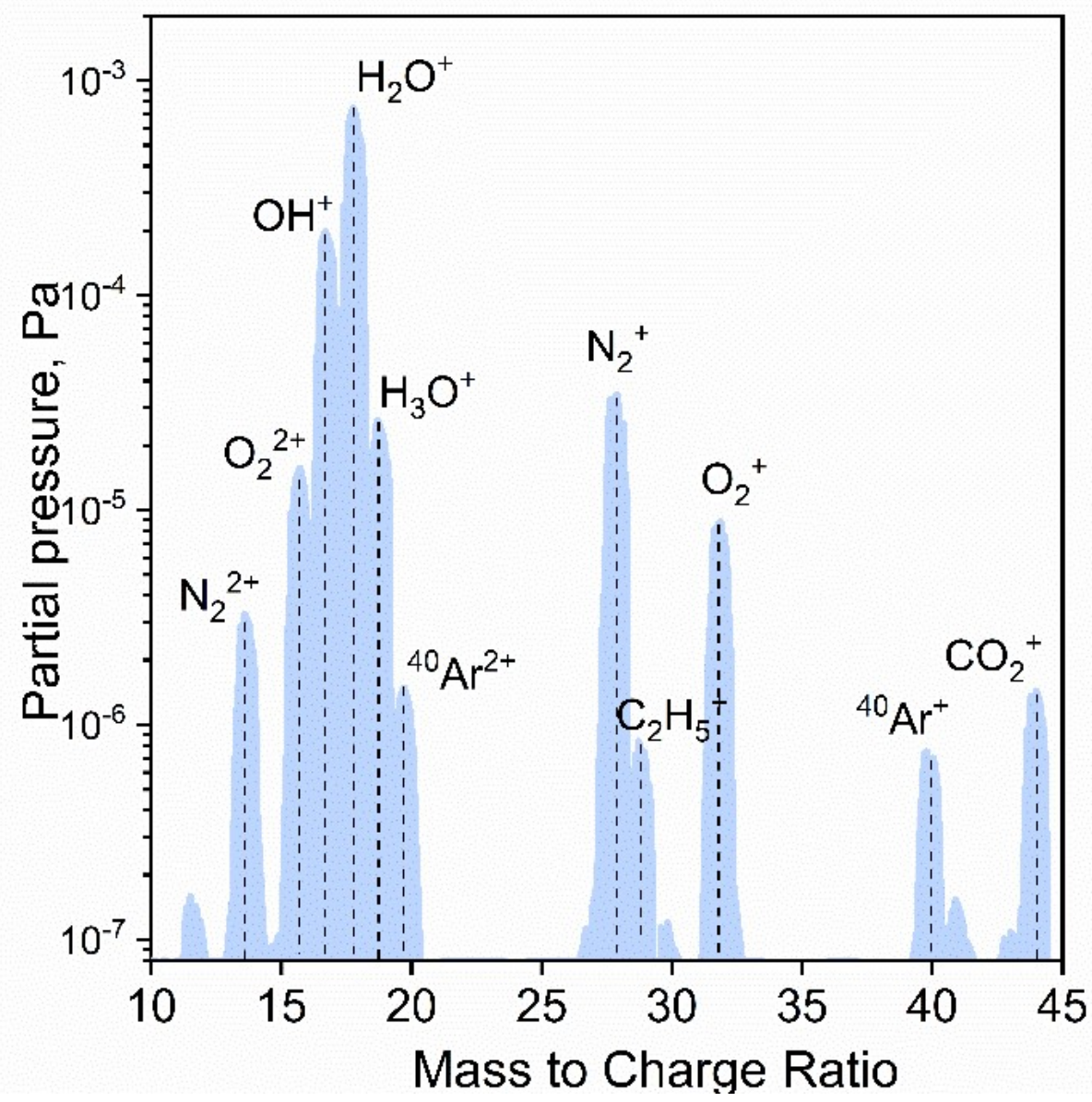

**Figure 1.** Residual gas composition in the sputtering chamber after argon purging, as measured by RGA.

Prior to deposition, the sputtering chamber was purged with argon and evacuated by vacuum pumps to a residual gas pressure of 4 × $10^{-3}$ Pa. Deposition was carried out at an argon working pressure of 0.2 Pa. High-purity argon (99.998%) was used throughout the sputtering process. Substrates were not heated during deposition. The composition of residual gases after argon purging of the chamber (immediately prior to deposition), as measured by an Extorr XT-100 residual gas analyzer (RGA), is presented in Fig. 1.

The magnetron power settings during deposition were as follows: for the Cu target, the power for most samples ranged between 180–220 W, with a current of 0.7–0.9 A (for the deposition of 2-nm-thick layers, the power was approximately 150 W at a current of 0.30 A); for the Ti target, the power ranged between 830–870 W, with a current of 1.8–2.0 A (for 2-nm-thick layers, the power was approximately 185 W at a current of 0.20 A). Power fluctuations during a single deposition run did not exceed 1%.

X-ray diffraction and reflection curves were obtained at a Bruker “D8 DISCOVER” diffractometer at the “Centre for X-ray Diffraction Studies”, the Research Park of Saint-Petersburg State University. It is equipped with a photon source operating at a constant wavelength of 1.5406 Å. All measurements were performed in geometry θ-2θ.

Transmission electron microscopy (TEM) measurements were carried out using Libra 200FE (Carl Zeiss, Germany) microscope (accelerating voltage 200 kV) at the “Interdisciplinary Resource Centre for Nanotechnology” Resource Center of Saint-Petersburg State University. This microscope is equipped with a highly efficient field emission emitter and an energy Ω filter to perform precision measurements in high-resolution mode. The thin cross-section foils for TEM investigations were fabricated by mechanical grinding with the subsequent polishing by low-energy $Ar^+$ ion irradiation. The electron diffraction patterns were obtained from selected areas with a diameter of 80 nm (SAED). The results of scanning TEM in high-angle annular dark-field (STEM-HAADF) mode were obtained with condenser aperture of 10 μm and with magnification of 150k. More detailed cross-sectional imaging was performed with high resolution TEM (HRTEM).

## Results and Discussion

The structural evolution of the $[Ti/Cu]_N$ multilayers was analyzed by combining local electron microscopy with integral X-ray methods. Since cross-sectional TEM analysis is experimentally demanding and requires complex and time-consuming preparation of individual thin lamellae, it was performed for representative multilayer structures rather than for the entire series, while XRD and XRR measurements were carried out for the entire series of synthesized samples. This approach makes it possible to correlate detailed local observations of layer morphology, crystallinity, and texture with the general trends in phase-related diffraction contributions and long-range periodicity as a function of multilayer period.

### 1. Transmission electron microscopy

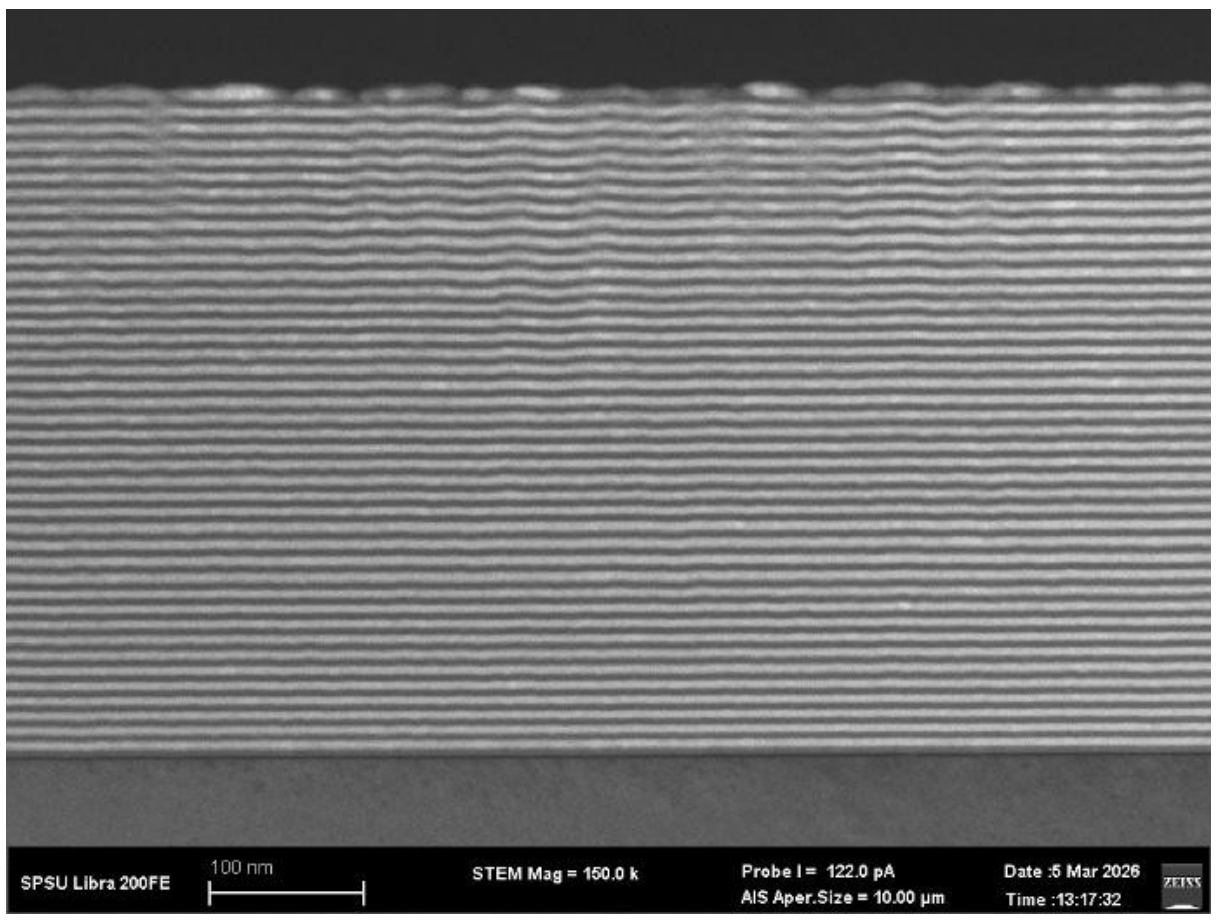

**Figure 2.** STEM-HAADF image of the cross-sectional view of the multilayer structure $[Ti/Cu]_{42}$ with a period of d = 10 nm.

Fig. 2 shows a STEM-HAADF image of the cross-section of the multilayer structure $[Ti/Cu]_{42}$ with a period of d = 10 nm. A periodic alternation of layers is clearly observed throughout the entire coating thickness, indicating that the layered structure is preserved during synthesis. The observed contrast is typical for the HAADF mode and is governed by the difference in the average atomic numbers of the elements: the Cu layers exhibit higher signal intensity, whereas the Ti layers appear darker.

The image analysis shows that the layers near the substrate are relatively smooth. However, as the distance from the substrate increases toward the top of the multilayer coating, the layers become progressively wavier. Despite this, the individual layers remain distinguishable, indicating that the periodic structure has not undergone complete degradation. The upper surface region differs from the rest of the multilayer stack. In this area, a discontinuous layer with an island-like morphology is observed, which can be attributed to a copper-based oxide phase formed as a result of oxidation of the top Cu layer upon exposure of the sample to the atmosphere.

The increase in layer waviness with the growth of the multilayer structure is commonly attributed to the accumulation of growth-related inhomogeneities, which, in turn, may be associated with changes in the structural state of the layers during deposition. To analyze the crystallinity of the multilayer system, SAED patterns and high-resolution TEM images were obtained from two regions of the sample: near the substrate and close to the surface. The corresponding results are presented in Figs 3(a, c) and 3(b, d).

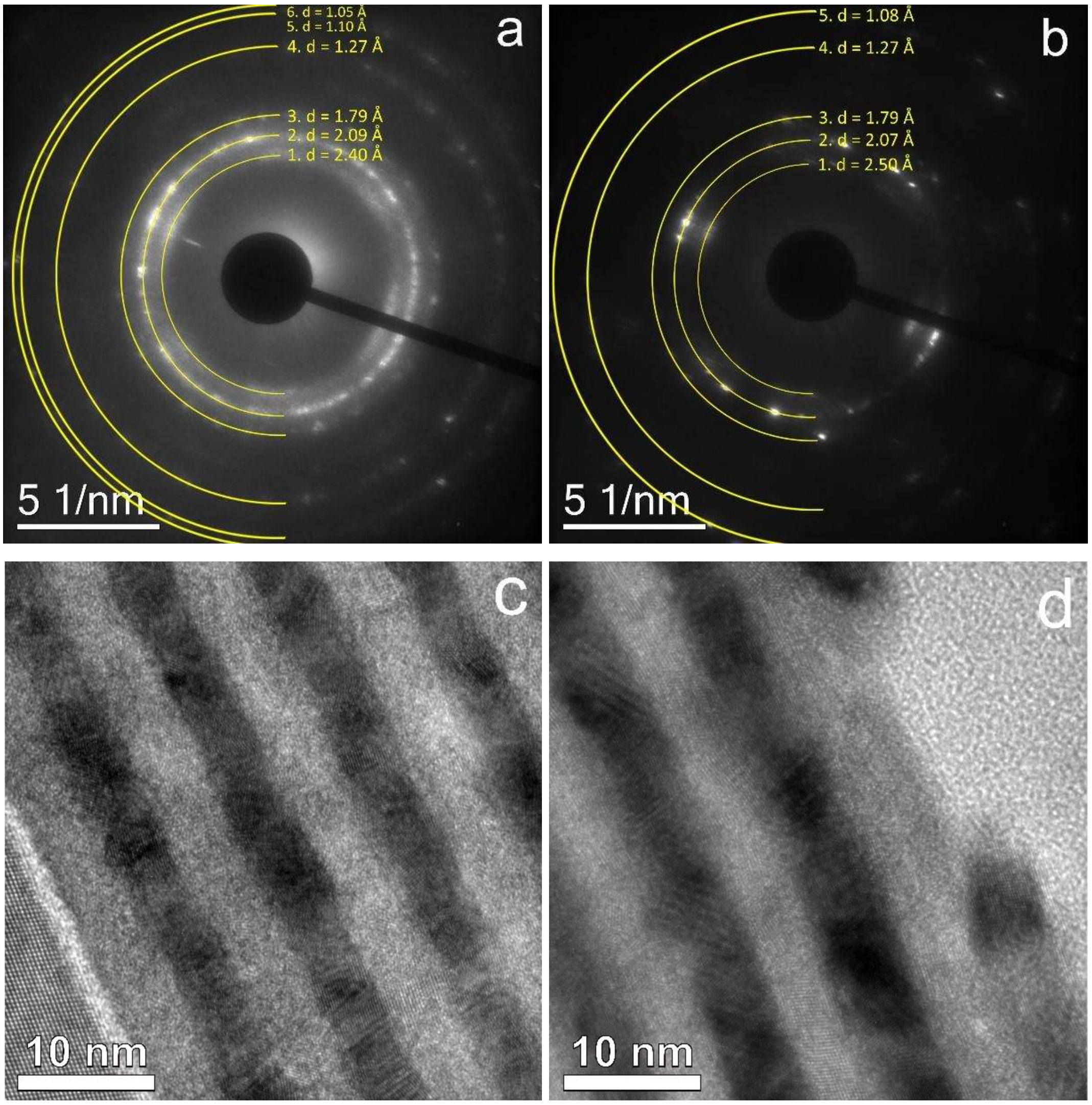

**Figure 3.** SAED (a, b) and HRTEM (c, d) images obtained for the multilayer structure $[Ti/Cu]_{42}$ with a period of d = 10 nm near the substrate (left) and close to the surface (right).

The SAED pattern obtained near the substrate (Fig. 3a) is characterized by the presence of bright but significantly broadened diffraction arcs. Such a shape of the reflections indicates that, at the initial stages of growth, the layers are in an amorphous-nanocrystalline or weakly crystallized state. At the same time, six arcs can be reliably distinguished in the diffraction pattern. In contrast, in the region near the surface (Fig. 3b), the arcs become narrower and locally split into bright azimuthally confined maxima. This indicates a more pronounced crystallization of the layers and the presence of texture. In this case, five main arcs can be clearly identified, which is most likely related not to a reduction in the number of reflections as such, but to their partial overlap and redistribution of intensity as the material becomes more structurally ordered.

Additional information on the local structure is provided by the HRTEM images (Figs 3c, 3d). In this case, the Cu layers appear darker compared to the Ti layers. In the region near the substrate, predominantly low-contrast and unevenly expressed lattice fringes are observed, which is consistent with a low degree of crystallinity of the Ti layers and the presence of only small nanocrystalline regions. At the same time, in the Cu layers, the periodic fringes are much more pronounced, indicating a higher local degree of ordering in Cu compared to Ti. In the region near the surface, both the Cu and Ti layers exhibit more extended areas with clearly resolved lattice fringes of different orientations, corresponding to a polycrystalline structure and consistent with the textured character of the SAED pattern. Thus, the HRTEM results confirm the SAED-based conclusion of a gradual transition from an amorphous-nanocrystalline state in the lower part of the coating to a more pronounced crystalline state in the upper part of the multilayer system.

For all observed arcs in the SAED images, the interplanar spacings d were determined and are summarized in Table 2. The obtained values unambiguously indicate the presence of copper crystallites: the corresponding reflections are in good agreement with tabulated values for fcc-Cu [13,14], and the difference between the experimental and reference d-spacings is minimal. In particular, the observed arcs can be attributed to reflections from the {111}, {200}, {220}, {311}, and {222} planes of fcc-Cu.

**Table 2.** Measured interplanar spacings $d_{meas}$ for the $[Ti/Cu]_{42}$ structure with a period of d = 10 nm (near the substrate / near the surface) and their interpretation based on literature $d_{lit}$ values.

| Arc No. | $d_{meas}$, Å | Phase | Planes | $d_{lit}$, Å |
|---|---|---|---|---|
| 1 | 2.40 / 2.50 | fcc-Ti / α-Ti | {111} / {100} | 2.37 / 2.55 |
| 2 | 2.09 / 2.07 | fcc-Cu | {111} | 2.09 |
| 3 | 1.79 / 1.79 | fcc-Cu | {200} | 1.80 |
| 4 | 1.27 / 1.27 | fcc-Cu | {220} | 1.27 |

| 5 | 1.10 / 1.08 | fcc-Cu | {311} / {311} + {222} | 1.09 |
|---|---|---|---|---|
| 6 | 1.05 | fcc-Cu | {222} | 1.04 |

The interpretation of the remaining arcs is less straightforward. For the region near the substrate, a d value of about 2.4 Å is observed, which is closest to the reflection from the {111} planes of fcc-Ti (d ≈ 2.37 Å [15,16]). In the near-surface region, the corresponding value increases to approximately 2.5 Å and becomes closer to the reflection from the {100} planes of α-Ti (d ≈ 2.55 Å [15,17]). Thus, the SAED analysis may indicate a change in the structural state of the titanium layers as the multilayer structure grows: while in the lower part of the sample Ti formation is closer to an fcc-like state, in the upper part the contribution of α-Ti becomes more probable. However, this conclusion should be considered preliminary, since the identification of titanium-related reflections is complicated by the limited number of reflections and the close values of interplanar spacings for the possible phases.

Thus, the SAED and HRTEM results consistently show that, with the growth of the $[Ti/Cu]_{42}$ multilayer structure with a period of d = 10 nm, the layers become more crystalline and more textured. This structural ordering during deposition is likely associated with the observed increase in layer waviness in the upper part of the coating. At the same time, morphological inhomogeneity of the top Cu layer is retained near the surface, and its subsequent oxidation in air apparently leads to the formation of a discontinuous oxide region with a “ragged” island-like morphology.

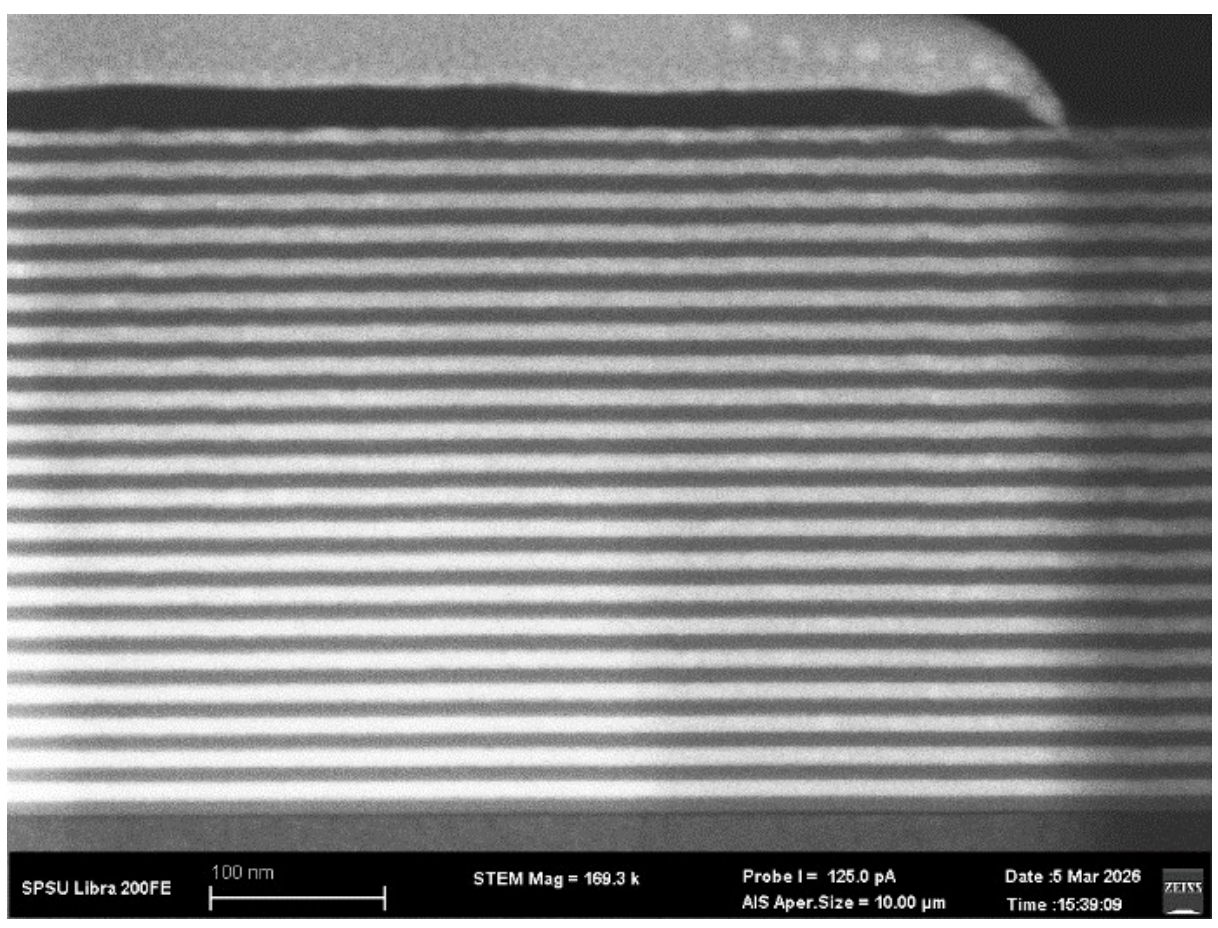

**Figure 4.** STEM-HAADF image of the cross-sectional view of the multilayer structure $[Ti/Cu]_{21}$ with a period of d = 20 nm.

Fig. 4 presents a STEM-HAADF image of the cross-section of the multilayer structure $[Ti/Cu]_{21}$ with a period of d = 20 nm. A regular alternation of layers is clearly observed throughout the entire coating thickness, indicating that the multilayer architecture is preserved during synthesis. In contrast to the system with a smaller period, this structure does not exhibit a pronounced increase in layer waviness with increasing coating thickness; the layered morphology remains more regular across the entire sample. In addition, the top Cu surface layer appears less inhomogeneous and less “ragged” than in the system with d = 10 nm, which may indicate a more stable morphology and, consequently, a less pronounced inhomogeneity of its subsequent oxidation in air.

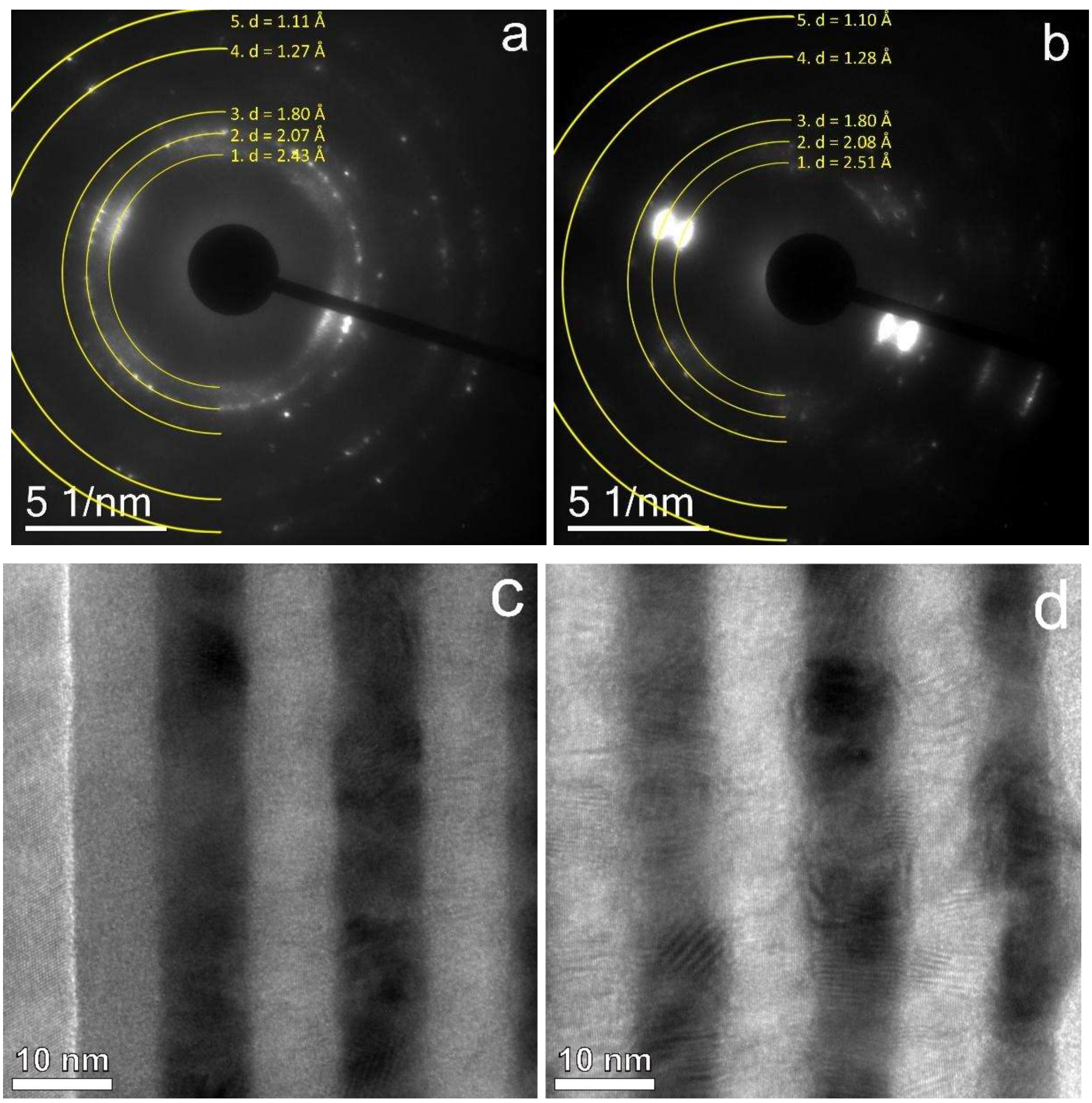

**Figure 5.** SAED (a, b) and HRTEM (c, d) images obtained for the multilayer structure $[Ti/Cu]_{21}$ with a period of d = 20 nm near the substrate (left) and close to the surface (right).

Figs 5a and 5b show SAED patterns obtained for the $[Ti/Cu]_{21}$ structure with a period of d = 20 nm near the substrate and near the surface, respectively. In both cases, five diffraction arcs can be clearly identified, and already in the region near the substrate the arcs contain azimuthally localized bright maxima. This indicates that even at the initial stages of growth, the layers are in a more crystalline state than in the system with $d$ = 10 nm and exhibit features of a textured polycrystalline structure. In the near-surface region, the azimuthal localization of intensity becomes even more pronounced: for arcs 1, 2, 4, and 5, the most intense reflections are concentrated within limited angular ranges. Such a redistribution of intensity indicates the development of a more pronounced texture as the coating grows.

Interplanar spacings d were determined from the SAED patterns and are presented in Table 3. As in the case of the structure with d = 10 nm, the d values corresponding to copper reflections are in good agreement with reference data for fcc-Cu. The observed arcs can be reliably assigned to reflections from the {111}, {200}, {220}, and {311} plane families, with the highest-angle reflection near the surface likely involving an overlap of contributions from the {311} and {222} planes. This indicates the formation of fcc-structured crystallites in the Cu layers throughout the entire coating thickness.

**Table 3.** Measured interplanar spacings $d_{meas}$ for the $[Ti/Cu]_{21}$ structure with a period of d = 20 nm (near the substrate / near the surface) and their interpretation based on literature $d_{lit}$ values.

| Arc No. | $d_{meas}$, Å | Phase | Planes | $d_{lit}$, Å |
|---|---|---|---|---|
| 1 | 2.43 / 2.51 | fcc-Ti / α-Ti | {111} / {100} | 2.37 / 2.55 |
| 2 | 2.07 / 2.08 | fcc-Cu | {111} | 2.09 |
| 3 | 1.80 / 1.80 | fcc-Cu | {200} | 1.80 |
| 4 | 1.27 / 1.28 | fcc-Cu | {220} | 1.27 |
| 5 | 1.11 / 1.10 | fcc-Cu | {311} / {311} + {222} | 1.09 |

The interpretation of reflections with d ≈ 2.43 Å near the substrate and d ≈ 2.51 Å near the surface is less straightforward. As in the case of the system with d = 10 nm, it is most likely that in the lower part of the coating these reflections are associated with an fcc-like state of titanium and can be attributed to reflections from the {111} planes of fcc-Ti, whereas in the upper part of the coating they are closer to reflections from the {100} planes of α-Ti. Thus, the SAED data also suggest that, as the coating grows, the structural state of Ti evolves from a metastable fcc-like state toward a state closer to α-Ti.
Additional insight into the local structure is provided by the HRTEM images (Figs 5c and 5d). Near the substrate, no clearly resolved lattice fringes are observed in the first Ti layer, which allows it to be characterized as amorphous or amorphous-nanocrystalline. In the subsequent Ti layers, signs of crystallization become more pronounced. At the same time, in the Cu layers, lattice fringes are already observed in the first layer adjacent to the substrate, indicating a higher tendency of copper to crystallize compared to titanium. In the near-surface region, both Cu and Ti layers exhibit more extended areas with well-defined lattice fringes, and their preferential orientation along the plane of the layers is consistent with the presence of texture revealed by SAED.
Overall, the SAED and HRTEM results show that in the $[Ti/Cu]_{21}$ multilayer structure with a period of d = 20 nm, crystallization develops already at the early stages of growth and remains pronounced throughout the entire coating thickness. Compared to the system with d = 10 nm, this structure is characterized by lower layer waviness and a more regular morphology, which correlates with a more stable development of a textured crystalline state during deposition. It is possible that the earlier formation of crystallographic domains oriented parallel to the layers contributes to the stabilization of layer growth and reduces the tendency for waviness to increase with thickness.

## 2. X-ray diffraction and reflectometry

Fig. 6 presents X-ray diffraction patterns of $[Ti/Cu]_N$ multilayer structures with periods of 4, 10, 15, 20, 30, and 52.5 nm, as well as reference Cu and Ti films with a thickness of 100 nm. The measurements were performed in the 2θ angular range from 25° to 57°. In all diffraction patterns, the reflection from the Si(111) substrate is observed as a peak at 2θ ≈ 28.4°.

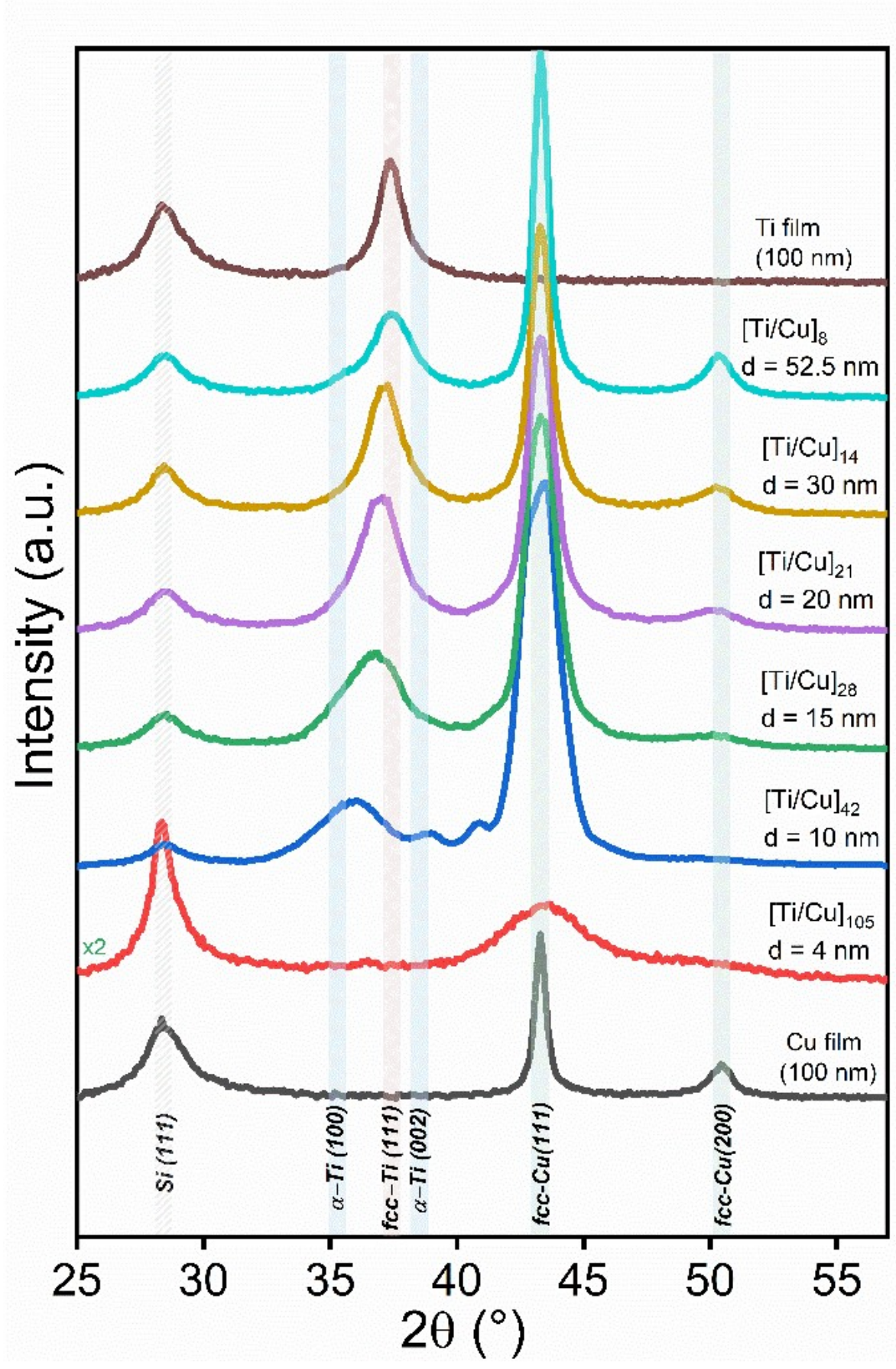

**Figure 6.** X-ray diffraction patterns of $[Ti/Cu]_N$ multilayer structures with periods of d = 4, 10, 15, 20, 30, and 52.5 nm, as well as 100 nm thick Cu and Ti reference films.

In the diffraction pattern of the reference titanium film, a single pronounced peak is observed at 2θ ≈ 37.4°. In thin titanium films deposited by magnetron sputtering, a reflection in this region is often associated with a preferred α-Ti(002) texture. This reflection is observed independently of the substrate onto which the film is deposited [18–23]. Its position may deviate from the standard value for α-Ti(002) (38.4°, JCPDS-ICDD card No. 44-1294) due to residual stresses, defects, impurity content, or specific deposition conditions [19]. At the same time, in a number of studies [2,3,24], this peak is interpreted as the (111) reflection of a metastable fcc-Ti phase (typically for Ti films up to 220 nm thick). Several possible mechanisms for the stabilization of this phase are discussed in the literature:

- The contribution of surface and interfacial energy becomes comparable to the bulk free energy ΔG [4,17]. However, first-principles (DFT) calculations indicate that the dominant fcc phase is stable only for film thicknesses up to ~20 nm.
- Rapid relaxation of adatoms at low substrate temperatures (<150 °C) and high deposition rates (>0.2 nm/s) leads to the formation of stochastic stacking faults, which cooperatively give rise to macroscopic fcc domains without transformation into the stable hcp structure [2,24].
- The emergence of biaxial stresses due to epitaxial matching with the substrate [25,26].

It should be emphasized that an unambiguous separation of the contributions from α-Ti(002) and fcc-Ti(111) based solely on the position of this peak is difficult, since the corresponding interplanar spacings are close. Therefore, this peak is hereafter considered as a Ti-related contribution, which may correspond either to textured regions of α-Ti with an orientation close to (002) or to a metastable fcc-like state of Ti with (111) orientation.

In the diffraction patterns of $[Ti/Cu]_N$ multilayer structures with periods of 10–52.5 nm, an asymmetric diffraction peak is observed in the $2\theta \approx 35°–37°$ range, whose shape, full width at half maximum, and position systematically change with the period of the structure. This profile can be described as a superposition of at least two reflections: a low-angle component close to the α-Ti(100) reflection ($2\theta \approx 35.1°–35.4°$ [15,17]) and a higher-angle component that may correspond to α-Ti(002) and/or fcc-Ti(111), as discussed above. With increasing multilayer period, the center of gravity of the peak shifts from lower angles toward a position characteristic of the reference titanium film, reflecting a redistribution of intensity between its constituent components. It can be concluded that in the multilayer structure with a period of 10 nm, the α-Ti(100) reflection predominates, whereas in samples with larger periods, reflections corresponding to other orientation(s) become dominant. For the multilayer with the smallest period of 4 nm, no peaks indicative of crystalline titanium phases are detected, suggesting that the Ti layers are amorphous at this scale of periodicity.

A combined analysis of the diffraction patterns of all studied multilayer structures and the 100 nm thick reference copper film reveals the presence of an intense reflection at $2\theta \approx 43.3°$ and a low-intensity peak in the region of $2\theta \approx 50.4°$. These maxima correspond to reflections from the (111) and (200) planes of the face-centered cubic lattice of copper, respectively [13,14]. The full width at half maximum of these peaks systematically decreases with increasing multilayer period, indicating an increase in the size of coherent scattering regions (CSRs) in the copper layers.

Particular attention should be paid to the appearance of low-intensity features at $2\theta \approx 38.9°$ and 40.8° in the diffraction pattern of the structure with a period of 10 nm (Fig. 7). One possible explanation for the origin of these peaks is the formation of Cu–Ti intermetallic phases. However, the periodic arrangement of these peaks relative to the main Cu(111) reflection and their small full width at half maximum, indicating a coherently scattering domain size exceeding the thickness of individual layers, support an alternative interpretation.

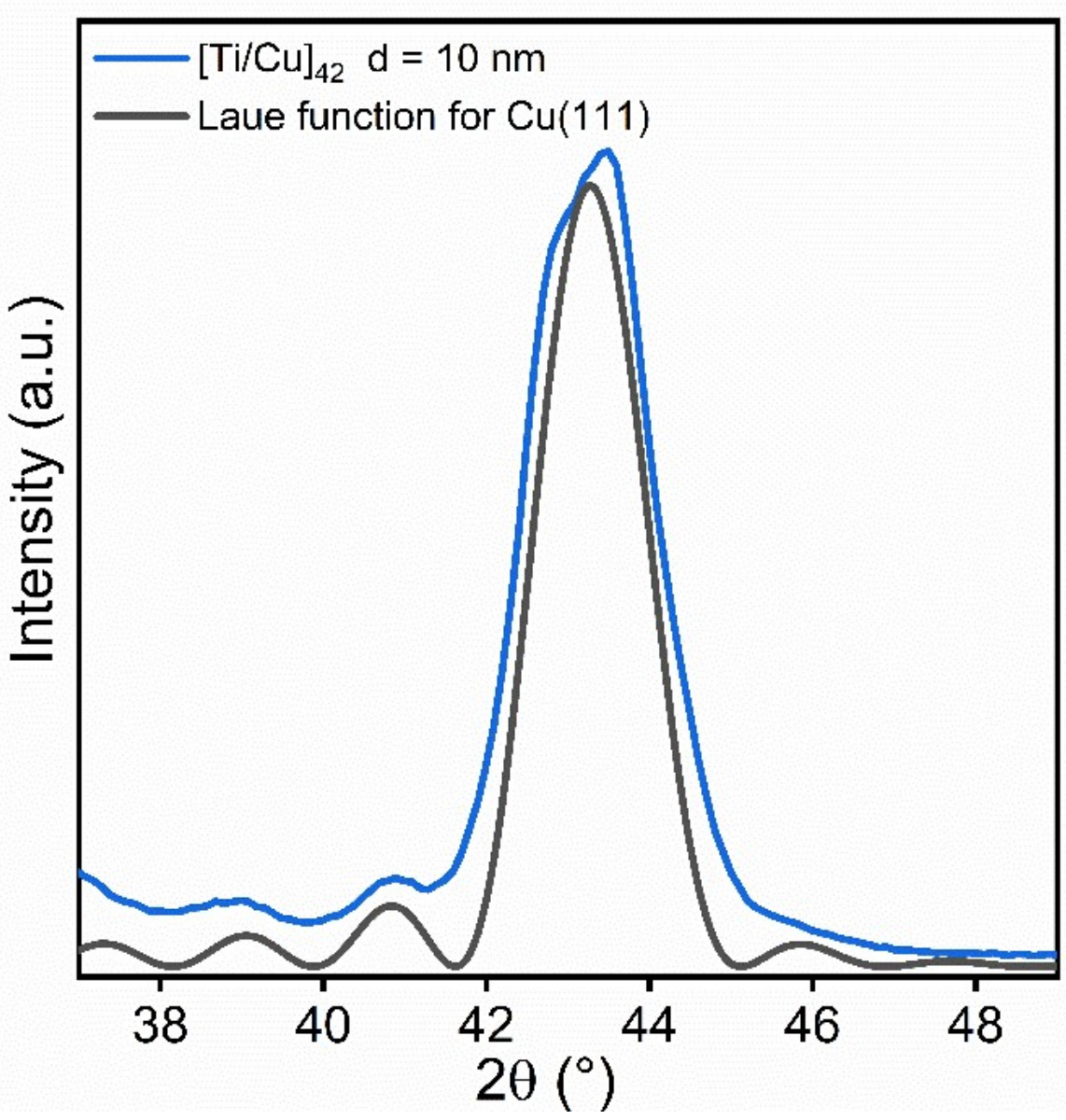

**Figure 7.** X-ray diffraction pattern of the $[Ti/Cu]_{42}$ multilayer structure with a period of 10 nm and the calculated Laue function: $\alpha = -3$, $N = 26$, $l = 2.085$ Å.

The presence of such peaks may be associated with Laue oscillations arising in crystallites of finite size with a narrow thickness distribution along the direction perpendicular to the reflecting atomic planes, provided that the multilayer structure exhibits high periodicity [27,28]. The intensity of such satellites can be described by the Laue function:

$$F(\theta) \propto \left(\frac{\sin(Nql/2)}{\sin(ql/2)}\right)^2 \tag{1}$$

$$q = \frac{4\pi}{\lambda}\sin\theta \tag{2}$$

where $l$ is the interplanar spacing for the crystallites; $\lambda$ is the X-ray wavelength (in our case 1.5406 Å); $N$ represents the number of spacings between atomic planes within a crystallite. In other words, the product $N \cdot l$ provides an estimate of the crystallite size along the direction perpendicular to the reflecting atomic planes, i.e., along the normal to the substrate surface in the θ–2θ geometry.

The asymmetry of the Laue satellite intensities and the presence of a low-angle "shoulder" of the main Cu(111) reflection in the diffraction patterns of the $[Ti/Cu]_{42}$ multilayer structure with a period of 10 nm indicates a non-uniform interplanar spacing $d$ along the growth direction. In the region of $2\theta \approx 42.5°$, a shift of intensity relative to the theoretical position of the bulk Cu reflection ($2\theta \approx 43.35°$ for $\lambda = 1.5406$ Å) is observed, corresponding to an effective interplanar spacing of $d \approx 2.125$ Å, which is ~1.9% larger than the tabulated value $d_{111} = 2.085$ Å.

For a quantitative description of the observed asymmetry of the satellite intensities, the standard Laue function is modified by introducing an exponential factor:

$$F(\theta) \propto \left(\frac{\sin(Nql/2)}{\sin(ql/2)}\right)^2 \cdot e^{\alpha(q_0-q)} \quad (3)$$

where α is the asymmetry parameter, and $q_0$ is the position of the central peak.

The plausible physical origin of the d-spacing gradient is the formation of a nonequilibrium solid solution of titanium in copper in the interfacial regions. Since Ti has a larger atomic radius than Cu ($r_{Ti} \approx 1.47$ Å, $r_{Cu} \approx 1.28$ Å), the incorporation of Ti atoms into the face-centered cubic lattice of Cu would be expected to expand the local lattice parameter, qualitatively consistent with Vegard-type behavior [29,30]. This scenario can explain both the shift of part of the Cu(111) intensity toward lower angles and the enhanced intensity of the low-angle Laue satellites: Cu regions adjacent to Ti-rich interfaces may therefore have larger interplanar spacings than the less intermixed parts of the crystallites.

An alternative, though not mutually exclusive, mechanism for the formation of the gradient may be the presence of coherent tensile strain. In multilayer nanostructures, macrostresses arising from lattice parameter mismatch or differences in thermal expansion coefficients of the components can lead to biaxial compression of copper within the plane of the layers and, consequently, to tensile deformation along the normal to the substrate. This also increases $d_{111}$ and shifts the scattering toward lower angles.

The contribution of metastable intermetallic phases in the Cu–Ti system (e.g., $TiCu_4$, whose reflections may fall within the 42°–43° range [31,32]) can be considered secondary, since the diffraction patterns do not exhibit the full set of characteristic peaks for these phases. Moreover, if such phases were formed predominantly in narrow interfacial regions, the size of the corresponding crystallites would be limited by the thickness of the transition layer; consequently, their reflections would be significantly broadened.

For a detailed comparison of the interplanar spacings obtained by X-ray diffraction (XRD) and selected area electron diffraction (SAED), peak deconvolution of the diffraction patterns was performed for multilayer structures with periods of 10 nm and 20 nm. During data processing, the background contribution was approximated by an exponential function, while the diffraction peaks were fitted using Gaussian functions.

Fig. 8 shows an example of the diffraction pattern decomposition for the Ti/Cu structure with a period of 20 nm. The satellites at the base of the Cu(111) reflection were fitted with symmetric functions, which is consistent with their interpretation as oscillations originating from the periodic structure. The asymmetric peak in the 2θ ≈ 35°–37° region is described as a superposition of two Gaussian components, in agreement with the previously proposed assumption of two structural contributions (α-Ti(100) with α-Ti(002) and/or fcc-Ti(111)).

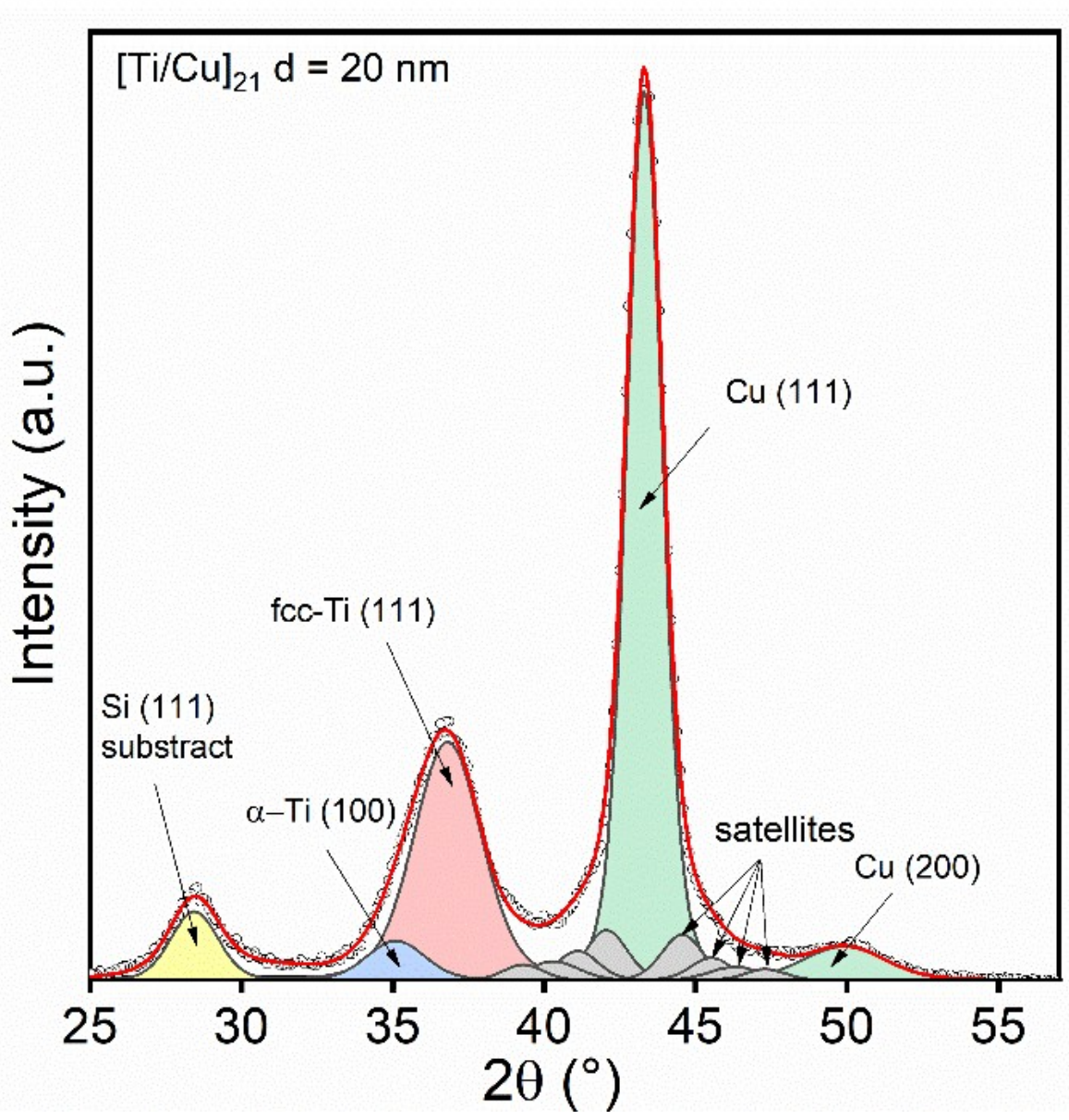

**Figure 8.** X-ray diffraction pattern of the $[Ti/Cu]_{21}$ multilayer structure with a period of 20 nm, decomposed into components.

The peak positions and the calculated interplanar spacings for the multilayer structures with periods of 10 nm and 20 nm are summarized in Table 4. For Cu(111), the obtained d values are about 2.08–2.09 Å, which is in good agreement with the tabulated value for fcc-Cu. The values corresponding to Cu (200) are also close to the expected ones for fcc-Cu, although this contribution is significantly weaker in the multilayer structures. Thus, the XRD results confirm the formation of predominantly (111)-textured fcc-Cu crystallites in the copper layers.

For the region with contributions from titanium, the peak decomposition yields two characteristic interplanar spacings: d ≈ 2.45–2.47 Å and d ≈ 2.55–2.56 Å. The first contribution is best associated with fcc-Ti(111) (for α-Ti(002), d ≈ 2.34 Å, whereas for fcc-Ti d ≈ 2.37 Å, which is closer to the obtained range), while the second is closer to the α-Ti(100) reflection. Such a separation is in good agreement with the SAED data: for the structure with d = 10 nm, a value of d ≈ 2.40 Å was observed in the lower part of the coating, close to fcc-Ti(111), whereas near the surface the value increased to d ≈ 2.50 Å and became closer to α-Ti(100). A similar trend was identified for the structure with d = 20 nm: near the substrate d ≈ 2.43 Å, while in the near-surface region d ≈ 2.51 Å. Thus, XRD captures the integral contribution of several structural states of Ti, whereas SAED and HRTEM allow these states to be correlated with the structural evolution across the thickness of the coating.

**Table 4.** Interplanar spacings $d_{meas}$ determined from the fitted peak positions after decomposition of the XRD patterns for the $[Ti/Cu]_N$ structures with a period of d = 10 nm and 20 nm.

| Peak | d=10 nm | | d=20 nm | |
|---|---|---|---|---|
| | 2θ | $d_{meas}$, Å | 2θ | $d_{meas}$, Å |

| fcc-Cu(200) | 49.3 | 1.85 | 49.9 | 1.83 |
|---|---|---|---|---|
| fcc-Cu(111) | 43.4 | 2.08 | 43.3 | 2.09 |
| fcc-Ti (111) | 36.3 | 2.47 | 36.8 | 2.44 |
| α-Ti (100) | 35.1 | 2.55 | 35.1 | 2.55 |

To further assess the quality of the periodic structure, X-ray reflectivity curves of $[Ti/Cu]_N$ multilayer coatings with different periods were measured, as shown in Fig. 9. The curves are vertically offset for clarity. In all cases, the positions of the main maxima shift toward lower angles with increasing multilayer period, and the total number of peaks within the range increases, which corresponds to the expected behavior of a multilayer structure with increasing period. At the same time, the shape and intensity of the peaks strongly depend on the period, reflecting differences in the degree of periodicity preservation and the quality of the interfacial boundaries.

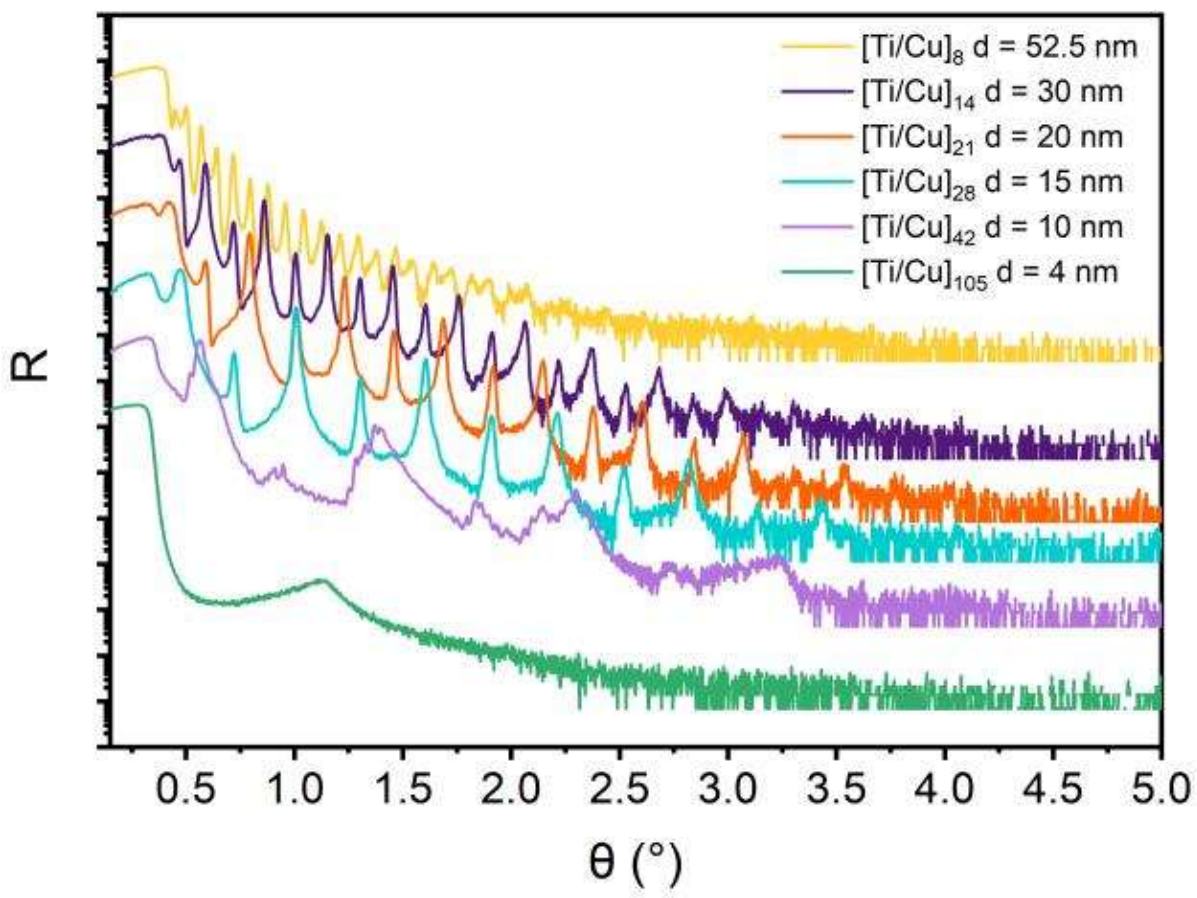

**Figure 9.** X-ray reflectivity curves of $[Ti/Cu]_N$ multilayer structures with nominal periods of d = 4, 10, 15, 20, 30, and 52.5 nm.

For the structure with the smallest period, d = 4 nm, the reflectivity curve contains only a weak and strongly broadened maximum. Such a shape indicates a low periodic contrast between the layers and/or significant smoothing of the chemical profile on the scale of the period. Possible reasons include the presence of extended interfacial regions, partial intermixing of Cu and Ti at the interfaces, and suppression of the formation of well-defined layers at very small individual layer thicknesses.

For the structure with d = 10 nm, Bragg maxima are already clearly observed, indicating that the periodic multilayer architecture is preserved. However, the peaks exhibit noticeable asymmetry and are elongated toward higher angles. This feature may indicate a distribution of periods across the coating thickness, as well as contributions from growth roughness and correlated morphological inhomogeneities. This interpretation is in good agreement with the STEM-HAADF results, which showed an increase in layer waviness with coating growth for the d = 10 nm structure.

With a further increase in the period to 15–52.5 nm, the Bragg peaks become more intense and better resolved, and their asymmetry becomes less pronounced. Qualitatively, this indicates a higher degree of periodicity in the multilayer structures and a reduced influence of accumulated growth roughness compared to the d = 10 nm structure.

Thus, the XRR data confirm the general trend revealed by TEM and XRD: at the minimum period of d = 4 nm, the multilayer structure is characterized by weak periodic contrast and, most likely, a significant contribution from interfacial intermixing; at d = 10 nm, periodicity is already preserved, but growth roughness and layer waviness have a noticeable effect on the shape of the Bragg maxima; at d ≥ 15 nm, more regular multilayer structures with well-defined periodicity are formed.

**Conclusion**

The structural evolution of $[Ti/Cu]_N$ multilayers was investigated using complementary electron microscopy, X-ray diffraction and reflectometry methods. The results obtained by these methods are mutually consistent and provide a coherent picture of the period-dependent transformation of the multilayer structure.

The present study demonstrates that the structural state of $[Ti/Cu]_N$ multilayers is governed by the competition between interfacial transition-region formation, crystallization, and the accumulation of growth-induced morphological instabilities. As the multilayer period increases from 4 to 52.5 nm, the system undergoes a transition from weakly contrasted nanoscale layering to well-defined periodic structures with pronounced crystalline ordering and texture.

At the smallest period of 4 nm, the periodic contrast is strongly reduced. This behavior may result from several mutually related factors, including extended transition regions at the interfaces, partial Cu–Ti intermixing, smoothing of the chemical profile over a length scale comparable to the bilayer period, and suppression of well-defined layer formation at very small individual layer thicknesses.

Increasing the period to 10 nm is sufficient to preserve the multilayer architecture, but this regime remains strongly affected by growth-induced morphological instabilities. Crystallization develops during growth, while accumulated roughness and layer waviness distort the long-range periodicity. At the same time, interlayer interaction remains significant on the scale of the bilayer period and may involve the formation of mixed Cu–Ti interfacial regions and/or coherent strain, both of which can locally modify the Cu lattice.

A further increase in the multilayer period leads to a qualitatively different growth regime. For periods of d ≥ 15 nm, the layers become sufficiently thick for more stable crystallization and texture development, while the periodic architecture is better preserved across the coating thickness. In this regime, the formation of crystallographic domains oriented predominantly along the layer plane appears to stabilize the multilayer growth and suppress the accumulation of waviness. The Cu layers crystallize predominantly in the fcc structure with a strong Cu(111) contribution, whereas the Ti layers exhibit a more complex structural evolution that cannot be reduced to a single unambiguous phase assignment. Instead, the Ti-related diffraction features indicate a gradual change in the local structural state of titanium as the coating grows.

These findings provide a structural basis for optimizing Ti/Cu multilayer coatings for applications requiring controlled interface sharpness, stable periodicity, and predictable diffraction or reflective performance.

**Acknowledgements**

This work was supported by the Russian Science Foundation (RSF), grant No. 24-72-10107. The authors gratefully acknowledge the staff of the "Centre for X-ray Diffraction Studies" and the "Interdisciplinary Resource Centre for Nanotechnology" of the Research Park of Saint Petersburg State University for their assistance with XRD, XRR, HRTEM, SAED, and STEM-HAADF measurements. The authors also thank the Petersburg Nuclear Physics Institute of the National Research Centre "Kurchatov Institute" (PNPI, NRC Kurchatov Institute) for the synthesis of the multilayer samples.

**References**

[1] L.P. Yue, W.G. Yao, Z.Z. Qi, Y.Z. He, Structure of nanometer-size crystalline Ti film, Nanostructured Materials 4 (1994) 451–456. https://doi.org/10.1016/0965-9773(94)90116-3.

[2] J. Chakraborty, K. Kumar, R. Ranjan, S.G. Chowdhury, S.R. Singh, Thickness-dependent fcc-hcp phase transformation in polycrystalline titanium thin films, Acta Mater. 59 (2011) 2615–2623. https://doi.org/10.1016/j.actamat.2010.12.046.

[3] M. Fazio, D. Vega, A. Kleiman, D. Colombo, L.M. Franco Arias, A. Márquez, Study of the structure of titanium thin films deposited with a vacuum arc as a function of the thickness, Thin Solid Films 593 (2015) 110–115. https://doi.org/10.1016/j.tsf.2015.09.015.

[4] R. Banerjee, S.A. Dregia, H.L. Fraser, Stability of f.c.c. titanium in titanium/aluminum multilayers, Acta Mater. 47 (1999) 4225–4231. https://doi.org/10.1016/S1359-6454(99)00281-5.

[5] J.M. Gómez-Guzmán, M. Opel, T. Veres, P. Link, L. Bottyán, Structural, electrical and magnetic properties of reactively DC sputtered Cu and Ti thin films. Application to Cu/Ti neutron supermirrors for low spin-flip applications, Nucl. Instrum. Methods Phys. Res. A 1059 (2024) 169005. https://doi.org/10.1016/j.nima.2023.169005.

[6] A. Hollering, N. Rebrova, C. Klauser, Th. Lauer, B. Märkisch, U. Schmidt, A non-depolarizing CuTi neutron supermirror guide for PERC, Nucl. Instrum. Methods Phys. Res. A 1032 (2022) 166634. https://doi.org/10.1016/j.nima.2022.166634.

[7] D. Dubbers, H. Abele, S. Baeßler, B. Märkisch, M. Schumann, T. Soldner, O. Zimmer, A clean, bright, and versatile source of neutron decay products, Nucl. Instrum. Methods Phys. Res. A 596 (2008) 238–247. https://doi.org/10.1016/j.nima.2008.07.157.

[8] J. Padiyath, J. Stahn, P. Allenspach, M. Horisberger, P. Böni, Influence of Mo in the Ni sublayers on the magnetization of Ni/Ti neutron supermirrors, Physica B Condens. Matter 350 (2004) E237–E240. https://doi.org/10.1016/J.PHYSB.2004.03.059.

[9] X. Wang, C. Ziener, H. Abele, S. Bodmaier, D. Dubbers, J. Erhart, A. Hollering, E. Jericha, J. Klenke, H. Fillunger, W. Heil, C. Klauser, G. Konrad, M. Lamparth, T. Lauer, M. Klopf, R. Maix, B. Märkisch, W. Mach, H. Mest, D. Moser, A. Pethoukov, L. Raffelt, N. Rebrova, C. Roick, H. Saul, U. Schmidt, T. Soldner, R. Virot, O. Zimmer, Design of the magnet system of the neutron decay facility PERC, EPJ Web Conf. 219 (2019) 04007. https://doi.org/10.1051/EPJCONF/201921904007.

[10] D. Dubbers, B. Märkisch, Precise Measurements of the Decay of Free Neutrons, Https://Doi.Org/10.1146/Annurev-Nucl-102419-043156 71 (2021) 139–163. https://doi.org/10.1146/ANNUREV-NUCL-102419-043156.

[11] N.K. Pleshanov, V.M. Pusenkov, A.F. Schebetov, B.G. Peskov, G.E. Shmelev, E. V. Siber, Z.N. Soroko, On the use of specular neutron reflection in the study of roughness and interdiffusion in thin-film structures, Physica B Condens. Matter 198 (1994) 27–32. https://doi.org/10.1016/0921-4526(94)90119-8.

[12] A. Torrisi, P. Horák, J. Vacík, V. Lavrentiev, A. Cannavò, G. Ceccio, J. Vaniš, R. Yatskiv, J. Grym, Synthesis of Cu–Ti thin film multilayers on silicon substrates, Bulletin of Materials Science 44 (2021) 1–8. https://doi.org/10.1007/S12034-020-02346-6/FIGURES/10.

[13] M.E. Straumanis, L.S. Yu, Lattice parameters, densities, expansion coefficients and perfection of structure of Cu and of Cu–In α phase, Acta Crystallographica Section A 25 (1969) 676–682. https://doi.org/10.1107/S0567739469001549.

[14] Zs. Czigány, F. Misják, O. Geszti, G. Radnóczi, Structure and phase formation in Cu–Mn alloy thin films deposited at room temperature, Acta Mater. 60 (2012) 7226–7231. https://doi.org/10.1016/j.actamat.2012.09.034.

[15] H.C. Wu, A. Kumar, J. Wang, X.F. Bi, C.N. Tomé, Z. Zhang, S.X. Mao, Rolling-induced Face Centered Cubic Titanium in Hexagonal Close Packed Titanium at Room Temperature, Sci. Rep. 6 (2016) 24370. https://doi.org/10.1038/srep24370.

[16] D.M. Tshwane, R. Modiba, A.S. Bolokang, Surface analysis of the stress-induced, impurity driven face centered cubic titanium phase and the ranging lattice parameter sizes, Mater. Today Commun. 24 (2020) 101168. https://doi.org/10.1016/j.mtcomm.2020.101168.

[17] J.X. Yang, H.L. Zhao, H.R. Gong, M. Song, Q.Q. Ren, Proposed mechanism of HCP → FCC phase transition in titianium through first principles calculation and experiments, Sci. Rep. 8 (2018) 1992. https://doi.org/10.1038/s41598-018-20257-9.

[18] V. Chawla, R. Jayaganthan, A.K. Chawla, R. Chandra, Microstructural characterizations of magnetron sputtered Ti films on glass substrate, J. Mater. Process. Technol. 209 (2009) 3444–3451. https://doi.org/10.1016/j.jmatprotec.2008.08.004.

[19] Y.L. Jeyachandran, B. Karunagaran, S.K. Narayandass, D. Mangalaraj, T.E. Jenkins, P.J. Martin, Properties of titanium thin films deposited by dc magnetron sputtering, Materials Science and Engineering: A 431 (2006) 277–284. https://doi.org/10.1016/j.msea.2006.06.020.

[20] A.Y. Chen, Y. Bu, Y.T. Tang, Y. Wang, F. Liu, X.F. Xie, J.F. Gu, Deposition-rate dependence of orientation growth and crystallization of Ti thin films prepared by magnetron sputtering, Thin Solid Films 574 (2015) 71–77. https://doi.org/10.1016/j.tsf.2014.10.053.

[21] M.J. Jung, K.H. Nam, L.R. Shaginyan, J.G. Han, Deposition of Ti thin film using the magnetron sputtering method, Thin Solid Films 435 (2003) 145–149. https://doi.org/10.1016/S0040-6090(03)00344-4.

[22] J.H. Kwon, D.Y. Kim, K.S. Kim, N.M. Hwang, Preparation of Highly (002) Oriented Ti Films on a Floating Si (100) Substrate by RF Magnetron Sputtering, Electronic Materials Letters 16 (2020) 14–21. https://doi.org/10.1007/s13391-019-00182-3.

[23] L. Skowronski, Optical properties of titanium in the regime of the limited light penetration, Materials 13 (2020). https://doi.org/10.3390/ma13040952.

[24] D. Dellasega, F. Mirani, D. Vavassori, C. Conti, M. Passoni, Role of energetic ions in the growth of fcc and ω crystalline phases in Ti films deposited by HiPIMS, Appl. Surf. Sci. 556 (2021) 149678. https://doi.org/10.1016/j.apsusc.2021.149678.

[25] A.A. Saleh, V. Shutthanandan, N.R. Shivaparan, R.J. Smith, T.T. Tran, S.A. Chambers, Epitaxial growth of fcc Ti films on Al(001) surfaces, Phys. Rev. B 56 (1997) 9841–9847. https://doi.org/10.1103/PhysRevB.56.9841.

[26] A.F. Jankowski, M.A. Wall, Formation of face-centered cubic titanium on a Ni single crystal and in Ni/Ti multilayers, J. Mater. Res. 9 (1994) 31–38. https://doi.org/10.1557/JMR.1994.0031.

[27] R.W.E. van de Kruijs, E. Zoethout, A.E. Yakshin, I. Nedelcu, E. Louis, H. Enkisch, G. Sipos, S. Müllender, F. Bijkerk, Nano-size crystallites in Mo/Si multilayer optics, Thin Solid Films 515 (2006) 430–433. https://doi.org/10.1016/j.tsf.2005.12.252.

[28] M.A. Hollanders, B.J. Thijsse, E.J. Mittemeijer, Amorphization along interfaces and grain boundaries in polycrystalline multilayers: An x-ray-diffraction study of Ni/Ti multilayers, Phys. Rev. B 42 (1990) 5481–5494. https://doi.org/10.1103/PhysRevB.42.5481.

[29] L. Vegard, VI. Results of crystal analysis, The London, Edinburgh, and Dublin Philosophical Magazine and Journal of Science 32 (1916) 65–96.

[30] A.R. Denton, N.W. Ashcroft, Vegard's law, Phys. Rev. A (Coll. Park). 43 (1991) 3161–3164. https://doi.org/10.1103/PhysRevA.43.3161.

[31] S. Semboshi, J. Ikeda, A. Iwase, T. Takasugi, S. Suzuki, Effect of Boron Doping on Cellular Discontinuous Precipitation for Age-Hardenable Cu–Ti Alloys, Materials 2015, Vol. 8, Pages 3467-3478 8 (2015) 3467–3478. https://doi.org/10.3390/MA8063467.

[32] S. Semboshi, S. Amano, J. Fu, A. Iwase, T. Takasugi, Kinetics and Equilibrium of Age-Induced Precipitation in Cu-4 At. Pct Ti Binary Alloy, Metallurgical and Materials Transactions A 2017 48:3 48 (2017) 1501–1511. https://doi.org/10.1007/S11661-016-3949-X.